\documentclass[fleqn,usenatbib]{mnras}

\usepackage[T1]{fontenc}

\DeclareRobustCommand{\VAN}[3]{#2}
\let\VANthebibliography\thebibliography
\def\thebibliography{\DeclareRobustCommand{\VAN}[3]{##3}\VANthebibliography}

\usepackage{graphicx}	
\usepackage{amsmath}	
\usepackage{amssymb}	
\usepackage{multirow}

\begin{document}

\label{firstpage}
\pagerange{\pageref{firstpage}--\pageref{lastpage}}

\title{A Dual-Zone Diffusion Model for High Energy Emissions of the Cygnus Cocoon}
\author[Zhan \& Wang]{Shihong Zhan$^{1,2}$, Wei Wang$^{1,2}$\thanks{E-mail: wangwei2017@whu.edu.cn}\\
$^{1}$Department of Astronomy, School of Physics and Technology, Wuhan University, Wuhan 430072, China\\
$^{2}$WHU-NAOC Joint Center for Astronomy, Wuhan University, Wuhan 430072, China \\
}
\maketitle
\begin{abstract}
{ As one of the brightest galactic ${\gamma}$-ray sources, the Cygnus Cocoon superbubble has been observed by many detectors, such as $Fermi$-LAT, ARGO, HAWC, and LHAASO. However, the origin of $\gamma$-ray emission for the Cygnus Cocoon and the possible contribution to PeV cosmic rays are still under debate. The recent ultrahigh-energy $\gamma$-ray observations by LHAASO up to 1.4 PeV towards the direction of the Cygnus Cocoon, as well as the neutrino event report of IceCube-201120A coming from the same direction, suggest that the Cygnus Cocoon may be one of the sources of high-energy cosmic rays in the Galaxy. In this work, we propose a dual-zone diffusion model for the Cygnus Cocoon: the cocoon region and surrounding interstellar medium (ISM). This scenario can account for the $\gamma$-ray data from GeV to $\sim$ 50 TeV and agree with the one sub-PeV neutrino event result from IceCube so far. Moreover, it predict a non-negligible contribution $\gamma$-ray emission at hundreds TeV from the ISM surrounding the Cygnus Cocoon. This possible diffuse TeV-PeV gamma-ray features can be resolved by the future LHAASO observations.   }
\end{abstract}


\section{Introduction}
Although it is well known that astrophysical sources in the Milky Way can accelerates cosmic rays (CRs) up to PeV, the counterpart of PeVatrons in the Milky Way is still ambiguous \citep{1984ARA&A..22..425H,1990acr..book.....B}. Commonly, supernova remnants (SNRs) are presumed to be the main source of very-high-energy cosmic rays in the Milky Way \citep{1934PNAS...20..259B,2004APh....21..241H}. However, theoretically, SNRs are difficult to get protons to PeV energies \citep{2013MNRAS.431..415B}, and there is no direct observational evidence to support the idea that SNRs can be PeVatrons \citep{2012SSRv..173..369H,2019APh...111..100A}. Another possible candidate is the galactic star-forming regions, where the overlapping stellar winds are anticipated to accelerate cosmic rays up to PeV energy levels \citep{1983SSRv...36..173C,2020SSRv..216...42B,2021MNRAS.504.6096M}. The powerful stellar wind acceleration model of Wolf-Rayet type stars may be an argument for the observed overabundance of $^{22} \rm Ne $ in galactic CRs \citep{2019JPhCS1400b2011K,2020MNRAS.493.3159G}. 

It is known that hadronic process can produce both high energy $\gamma$-rays and neutrinos. If the observation of neutrinos are associated with the site of ultrahigh-energy $\gamma$-rays, the object can be unambiguously identified as a Galactic CR PeVatrons. The IceCube collaboration reported a 154 TeV candidate neutrino event from the direction of the Cygnus region \citep{2020GCN.28927....1I,2021ApJ...916L..22D} through the standard BRONZE alert procedure \citep{2019ICRC...36.1021B}. \cite{2021ApJ...916L..22D} reported the observation of photons above 300 TeV, which can be coincident with the neutrino event temporally.     

Cygnus Cocoon, a superbubble surrounding the Cygnus OB2 region, is first identified by the Fermi Gamma-Ray Space Telescope’s Large Area Telescope (LAT) \citep{2011Sci...334.1103A}. According to $Fermi$-LAT, the energy distribution at GeV energies follow a power-law form with a spectral index 2.1. And the Cygnus Cocoon is estimated to be 1.4 kpc away from the earth. \cite{2014ApJ...790..152B} first identified the gamma-ray emission at TeV energies. \cite{2020PhRvL.124b1102A,2021NatAs...5..465A} reported the observations of 1-100 TeV $\gamma$-rays coming from the Cygnus Cocoon based on the High Altitude Water Cherenkov (HAWC) observations, which can be fitted by a power law model below 10 TeV and shows a spectral softening at 10 TeV. Recently, LHAASO, a TeV-PeV energy dual-task facility designed for cosmic-ray and $\gamma$-ray studies, reported the observation of very-hing-energy photon with up to 1.4 PeV from the Cygnus Cocoon direction, which indicate that the spectrum may extend up to $\sim$1 PeV \citep{2021Natur.594...33C,2022icrc.confE.843L}. Moreover, \cite{2021PhRvL.126n1101A} revealed that the Cygnus region has a significant contribution to the $\gamma$-ray emission above 400 TeV, which is compatible with expectations from the hadronic process. These findings suggest that the Cygnus Cocoon may be a Galactic hadronic PeVatron, which can help understanding the origin of the knee of the CR energy spectrum.

Whether the star-forming regions can indeed accelerate CRs can be verified by the spectral and spatial features of $\gamma$-ray. \cite{2019NatAs...3..561A} showed that the CRs density surrounding the OB-association Cygnus OB2 and compact clusters Westerlund 1 and Westerlund 2 behaves like $r^{-1}$ based on H.E.S.S. and $Ferimi$-LAT data. They concluded this is due to the constant injection of relativistic particles into the interstellar medium, which may be an evidence for combined action of multiple stellar winds. On the other hand, there thousands OB stars and low-mass young sources in the Cygnus OB2, which is near the Cygnus Cocoon \citep{2012A&A...539A..79R}. And the massive population of the Cygnus OB2 was estimated by \cite{2000A&A...360..539K} to about 2600 B stars and 120 O stars. The total mechanical power of the Cygnus Cocoon is over $10^{38}$ erg/s \citep{2019NatAs...3..561A}. These findings supports that the star-forming regions and massive stellar clusters are likely active particle acceleration region in Galactic, and also imply that the possible emission contribution of high energy particles escaping to the outside region of the Cygnus Cocoon.  

The observed 1-100 TeV $\gamma$-rays from the Cygnus Cocoon likely originate by hadronic process and are hardly explained by a single electron population emitting gamma rays by inverse-Compton process without its synchrotron radiation exceeding the flux constraints set by radio and X-ray data \citep{2021NatAs...5..465A}. The strong magnetic turbulence in the Cygnus Cocoon makes it harder for CRs to escape, then \cite{2011Sci...334.1103A} concluded that the diffusion length in the Cygnus Cocoon is about 100-1000 times shorter than that in the standard interstellar medium. Moreover, there is a break around a few TeV in gamma-ray spectrum, which could be caused either by the leakage of CRs from the Cocoon, or a cutoff in the injected comic ray spectrum from the source \citep{2021NatAs...5..465A}. These studies suggest that the diffusion process is very important to the distribution of high energy CRs in the Cygnus Cocoon. 

The effect of diffusion on the energy spectrum has been discussed \citep{2022ApJ...931L..30B}.  In this work, we further consider the natural phenomenon of particles spreading from the inside of the Cygnus Cocoon to ISM. Based on the above reasons, we estimate the $\gamma$-rays mainly produced by hadronic interaction and put forward a dual-zone diffusion model for the Cygnus Cocoon to explain the the gamma-ray spectrum from GeV to PeV, and predict the TeV-PeV diffuse radiation structure around the Cygnus Cocoon. In Section 2, we show the methodology and details of the model. The results of gamma-ray spectral fitting and the neutrino flux are shown in Section 3. Finally, we will discuss and make a conclusion in Section 4.

\section{Methodology}

According to the significance map of the Cocoon region \citep{2021NatAs...5..465A}, the angular size of the Cygnus Cocoon is around $2.1^{\circ}$, corresponding to a radius of $r_{c}=55\rm \, pc$. It is natural that high energy particles may escape from the acceleration region. In our work, we assume the region of CRs involved in the hadronic interaction can be broadly divided into two classes as illustrated in Fig. 1: the inside region of cocoon ( $r_c\sim 55 \rm \, pc$), and the interstellar medium (ISM) region from the Cygnus Cocoon (we will calculate the diffuse from the Cygnus Cocoon to the earth, $\sim 1.4 \rm \, kpc$).  High energy protons are injected in cocoon and the diffused. The two regions have different diffusion coefficients and gas density.

\begin{figure}
\begin{minipage}{0.50\textwidth}
 \includegraphics[width=0.88\textwidth]{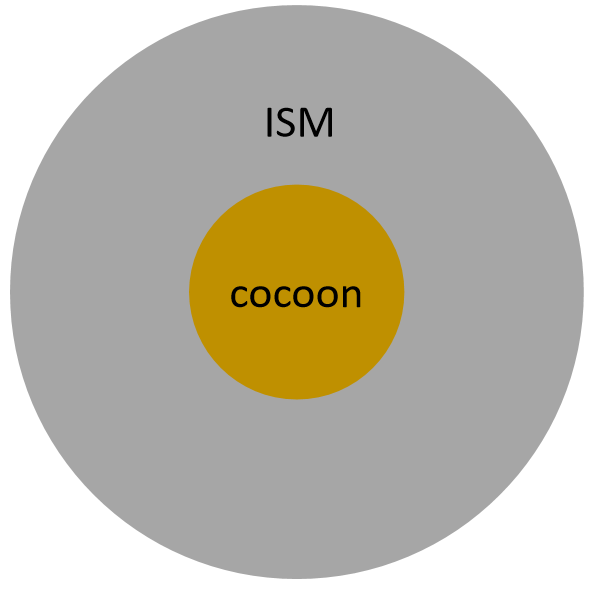}
\end{minipage}
\caption{The  schematic figure of particle transport in the Cygnus Cocoon. The yellow inner part represent the inside of the cocoon and the gray outer part represent the interstellar medium surrounding the cocoon. }
\label{fig1}
\end{figure}

There is a massive molecular cloud with a total mass of about $8\times 10 ^{6} $ $\rm M_{\odot}$ in Cygnus region \citep{2012A&A...538A..71A}, the gas number density should be more than 10 $\rm cm ^{-3}$ \citep{2014ApJ...790..152B}.  We take $n_{\rm c} = 30 \rm \, cm^{-3}$ in the Cygnus Cocoon, as suggested by the observations of \cite{2021NatAs...5..465A} and assume the number density of ISM is $1 \rm \, cm^{-3}$ . The age of the Cygnus OB2 is about $1-7\rm \, Myr$, we take $t_{age}=1 \rm \, Myr$ in our job. For simplicity, we consider a steady spherically symmetric case and assume that the injected particles follow a power-law energy distribution:
\begin{equation}
Q(E)=A\left(\frac{E}{1\rm \, GeV} \right)^{-\alpha}, \, \, 1.4\rm GeV < E < E_{max}
\end{equation}
where $A$ is the spectrum constant, $\alpha$ is the spectrum index, $E_{max}$ is the maximum energy of protons, $1.4 \rm \, GeV$ is the energy threshold for proton-proton(pp) interaction.

The accurate timescale calculation for a complex region like Cygnus Cocoon is difficult, but we can approximate the maximum energy range of CRs that may can be accelerated in Cygnus Cocoon by some simple discussion. At first, the acceleration timescale for particles by diffusive shock acceleration (DSA) mechanism with energy $E$ is (scaled to Bohm limit diffusion, \citealt{1983RPPh...46..973D}):
\begin{equation}
t_{\rm acc} \approx 1.1\times10^3 ~{\rm yr}\ \eta_{\rm acc} \eta_{\rm g} \frac{E}{1 \rm TeV} \left(\frac{B}{1\rm \mu G}\right)^{-1} \left(\frac{v}{1000 \, {\rm km~ s^{-1}}}\right)^{-2} ,
\end{equation}
where $\eta_{\rm acc}\leq 1$ accounts for the anisotropic scattering due to shock obliquity effect,  $\eta_{\rm g}\geq1$ accounts for the uncertainty of the particle diffusion relative to Bohm limit, $v$ is the velocity of shock. Here, we assume that the shock velocity is equal to the typical stellar wind $v_{w}=1000 \rm \, km \, s^{-1}$ \citep{2011Sci...334.1103A}. And we take $20\rm \mu G$ magnetic field which inferred from pressure balance with the gas in Cygnus Cocoon \citep{2011Sci...334.1103A}. For high energy ($E\geq1.4\rm \,GeV$) protons, the proton-proton inelastic collision timescale is given by \citep{1996A&A...309..917A}:
\begin{equation}
\tau_{\rm pp}\approx6\times10^{7}~{\rm yr}~\left(\frac{n_{\rm H}}{1~{\rm cm}^{-3}}\right)^{-1} . 
\end{equation}
where $n_{H}$ is the number density.  The particles should be confined in the shock to maintain acceleration, and the escaping boundary can be defined as a $\kappa r_{s}$ ahead of shock front and $\kappa=0.04-0.1$ \citep{2008ApJ...678..939Z,2011ApJ...731...87E}. The variable $r_{s}$ represents the radius of the shock responsible for accelerating particles to high energies within the Cygnus Cocoon. This shock can be regarded as the termination shock within the Cygnus Cocoon environment, which is approximately pc-order size. In our study, we adopt $r_{s}=1 \rm \, pc$. The time for particles to cross the boundary is given by:
\begin{equation}
\tau_{\rm esc}\approx 9\times 10^4 {\rm yr}  \eta_{\rm esc} \eta_{\rm g}^{-1}\left(\frac{E}{1\rm TeV}\right)^{-1} \left(\frac{B^f}{1\rm \mu G}\right) \left(\frac{r_{s}}{1\rm pc}\right)^2,
\end{equation}
where $\eta_{\rm esc}=\kappa^{2}\eta_{\rm acc}\leq 0.1$.  The parameter $B^{f}$ is the far upstream magnetic field which can be taken as $5 \rm \, \mu G$. With these formulas, we can get the maximum energy $\frac{E_{max}}{1 \rm \, TeV}=min(\frac{3.6\times 10^{4}}{\eta_{\rm acc} \eta_{\rm g}}, \frac{5\times 10^{3}}{\eta_{g}}\sqrt{\frac{\eta_{\rm esc}}{\eta_{\rm acc}}}, \frac{t_{age}}{5.5\rm \, yr})$. We show possible timescales for the relativistic protons in Cygnus Cocoon for two cases in Fig. 2, one corresponding to a maximum energy $E_{max}=100\rm \, TeV$ with characters $\eta_{\rm acc}=0.8$, $\eta_{\rm g}=1$,  $\eta_{\rm esc}=0.1$, and the other corresponding to a maximum energy $E_{max}=5 \rm \, PeV$ with characters $\eta_{\rm acc}=0.002$, $\eta_{\rm g}=1$,  $\eta_{\rm esc}=0.1$.

\begin{figure}
\begin{minipage}{0.50\textwidth}
 \includegraphics[width=0.88\textwidth]{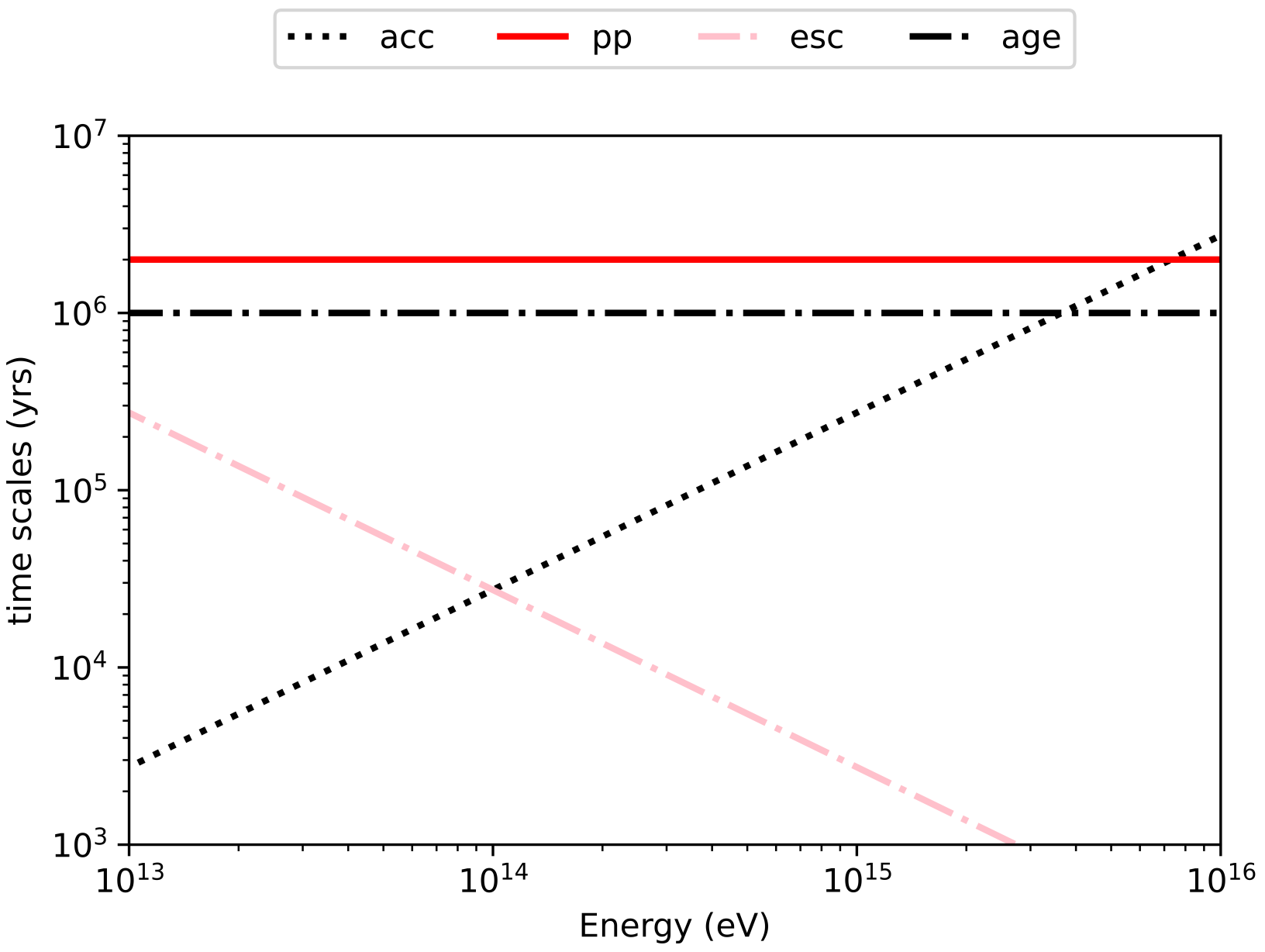}
  \includegraphics[width=0.88\textwidth]{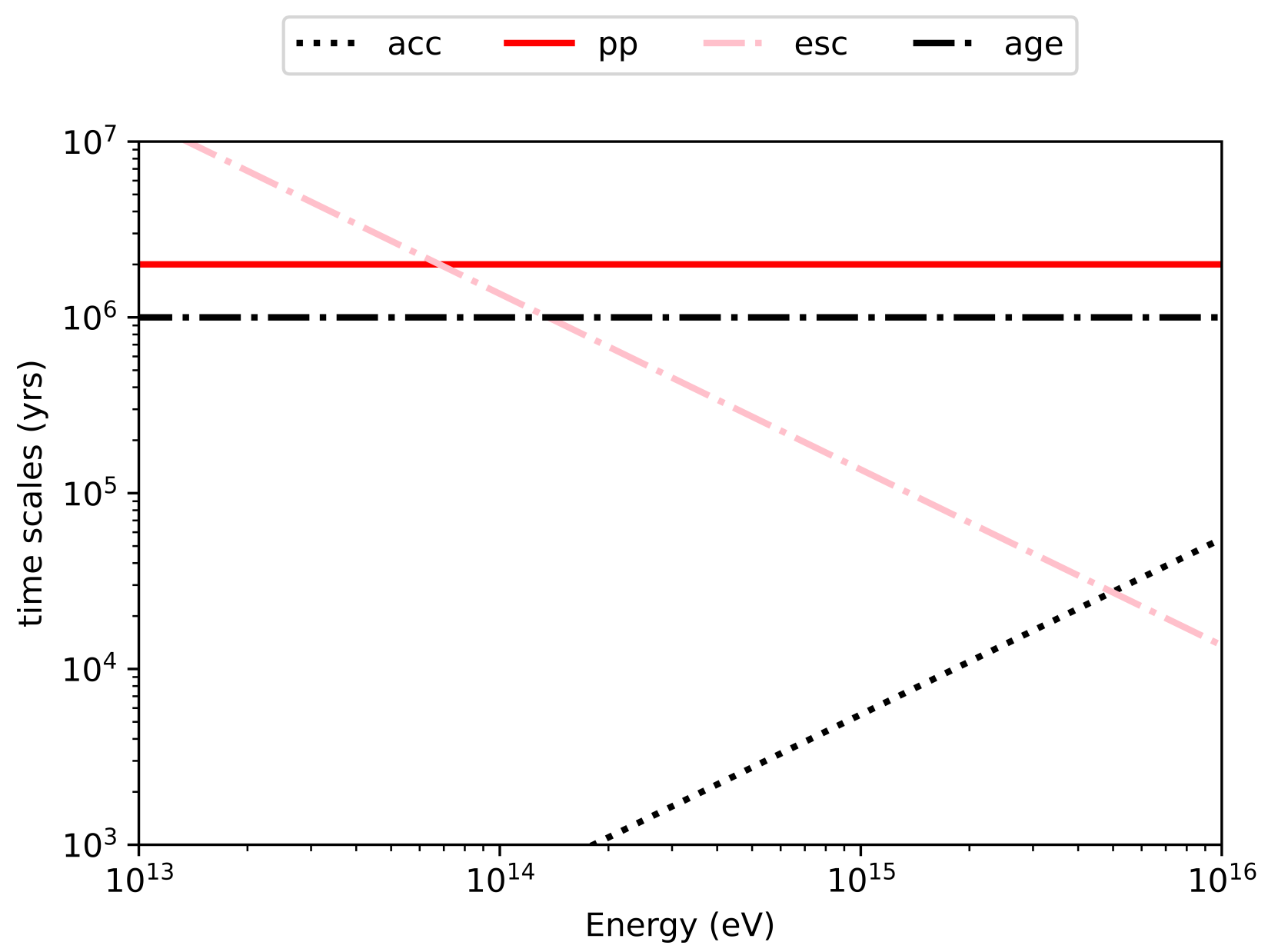}
\end{minipage}
\caption{{\em \rm Upper panel}: An example for timescales of relativistic protons in the Cygnus Cocoon with characters $\eta_{\rm acc}=0.8$, $\eta_{\rm g}=1$,  $\eta_{\rm esc}=0.1$. The curves represent the time scales of the acceleration (red solid line), pp energy loss (green dotted line), escaping (black solid line), and the age of the Cygnus Cocoon (black chain line). {\em \rm Bottom panel}: Possible timescales of relativistic protons for the different scenario with characters $\eta_{\rm acc}=0.002$, $\eta_{\rm g}=1$,  $\eta_{\rm esc}=0.1$.  }
\label{fig2}
\end{figure}

\section{Spectrum of Cygnus Cocoon}

The charged particle diffusion is caused by magnetic field scattering. The diffusion takes place in the resonant state, when the gyroradius of particles is equal to its wavelength \citep{2002cra..book.....S}. \cite{1990acr..book.....B} had presented the calculation method of the diffusion coefficient for particles in resonant state. For a single point source at $r_{s}$, the diffusion–convection equation of cosmic ray particles reads:

\begin{equation}
\begin{split}
    & \frac{\partial}{\partial t} n(E,r,t)-{\rm div}[D(E,r)\nabla n(E,r,t)]-\frac{\partial}{\partial E}[g(E)n(E,r,t)] = \\
    & Q(E,t)\delta^{3}(r-r_{s})
\end{split}
\end{equation}
where n(E,r,t) is the density of CR, $D(E,r)$ is the diffusion coefficient, $g(E) = -\dot{E}$ is the energy loss rate, and $Q(E,t)$ is the source function. We assume the diffusion coefficient $D(E,r)$ follow the form: 

\begin{equation}
D(E,r) = 
\begin{cases}
\ \left( \frac{E}{10\rm GeV} \right)^{\delta_{1}} D_{1},0<r<r_{c},\\
\ \left( \frac{E}{10\rm GeV} \right)^{\delta_{2}} D_{2},r\geq r_{c},
\end{cases}
\end{equation}
We assume the diffusion in ISM surrounding cocoon is the same with the interstellar diffusion of the Galaxy \citep{2011Sci...334.1103A}, which means $D_{2}=10^{28} \rm \, cm^{2}\, s^{-1}$, $\delta_{2}=0.5$, while the diffusion inside Cygnus Cocoon has two free parameters $D_{1}$ and $\delta_{1}$. The solution of Equ. (1) for a single-source spherically symmetric case has been presented by \cite{2019MNRAS.484.3491T}:

\begin{equation}
n(E,r) = \frac{1}{g(E)}\int_{max[0,t_{age}-\bar{t}_{c}(E_{max},E)]}^{t_{age}}g(E_{s})Q(E_{s},t_{s})H(r,E)dt_{s}
\end{equation}
where $E_{s}$ is the energy of the particle injected at $t_{s}$ from source, and we assume the injection of source is a constant injection in our job, which means $Q(E_{s},t_{s})=Q(E_{s})$. And $\bar{t}_{c}(E_{max},E)$ is the cooling timescale for a proton decaying from energy $E_{max}$ to energy $E$, which is defined as:

\begin{equation}
    \bar t_{c}(E_{max},E)=\int_{E}^{E_{max}}\frac{dE}{g(E)}
\end{equation}
where $g(E)$ is the energy loss rate, and $H(r,E)$ is defined as:

\begin{equation}
\begin{split}
H(r,E)  = \frac{b(b+1)} {\pi^{1.5}[2b^{2}erf(r_{c}) - b(b-1)  erf(2r_{c})+2erfc(r_{c}))]} \\
 \times  \begin{cases}
\ [e^{-r^{2}/r^{2}_{d1}}+M(\frac{2r_{c}}{r}-1)e^{-(r-2r_{c}^{2}/r_{d1}^{2})}],0<r<r_{c},\\
\ (1+M)[\frac{r_{c}}{r}+b(1-\frac{r_{c}}{r})]e^{-[(r-r_{c})/r_{d2}+r_{c}/r_{d1}]^{2}},r\geq r_{c},
\end{cases}
\end{split}
\end{equation}
where $M=\frac{b-1}{b+1}$, and $b$ is a constant can be defined as:

\begin{equation}
b=\sqrt{\frac{D(E,r<r_{b})}{D(E,r\geq r_{b})}}
\end{equation}
$r_{d1}$ and $r_{d2}$ are diffusion length scale which is defined as:

\begin{equation}
\begin{split}
r_{d1}(E) & =2\sqrt{\lambda (E,E_{s})} \\
& =2\sqrt{\int_{E}^{E_{s}}\frac{D(E,r<r_{b})}{g(E)}}
\end{split}
\end{equation}

\begin{equation}
\begin{split}
r_{d1}(E) & =2\sqrt{\lambda (E,E_{s})} \\
& =2\sqrt{\int_{E}^{E_{s}}\frac{D(E,r\geq r_{b})}{g(E)}}
\end{split}
\end{equation}
where $\lambda (E,E_{s})$ is the Syrovatsky variable \citep{1990A&A...232..582B}, which means the square of the distance travelled by a particle as its energy decays from $E_{s}$ to $E$.

In our work, we only consider the pp interaction for the energy loss of CRs. For simplicity, we use a simple assumption that the high energy protons lose half energy in every pp interaction and take the cross section as a constant $\sigma = 30 \rm mb$ to calculate the $g(E)$:
\begin{equation}
g(E)=Ec\sigma n_{H} \ln 2
\end{equation}
where $n_{H}$ is the background number density of hydrogen particles, $c$ is light speed. 

\subsection{Gamma-ray emission}

Gamma-ray hadronic mechanism emission due to neutral pions decay produced by pp interaction has long been studied from  high energy astronomy cases. We use the parameterizations based on \cite{2021PhRvD.104l3027K} to calculate the gamma-ray emission due to pp interaction. We have estimated the radiation from hadronic process dominates the gamma-ray band and the environment is spherically symmetric to use Equ. (5). 

The main parameters in our work are diffusion coefficient $D_{1}$, the spectral constant $A$, spectral index $\alpha$ and the maximum energy $E_{max}$. The data we used are taken from $Fermi$-LAT (\citep{2011Sci...334.1103A}), ARGO (\citep{2014ApJ...790..152B}), HAWC (\citep{2021NatAs...5..465A}). The resulting $\gamma$ -ray and the spectrum of protons are shown in Fig. 3, the relevant parameters are shown in Table 1. 

\begin{figure}
\begin{minipage}{0.50\textwidth}
 \includegraphics[width=0.9\textwidth]{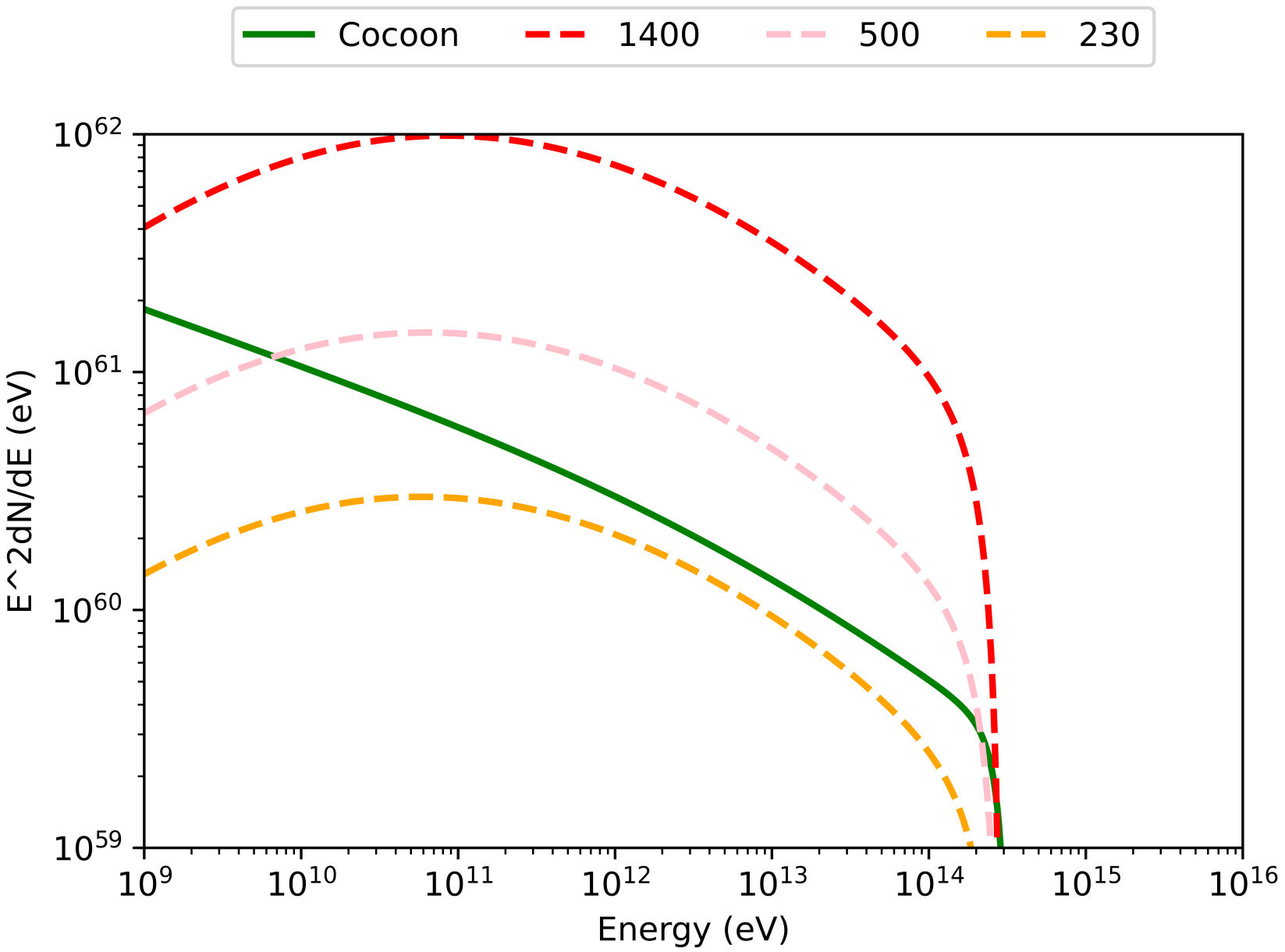}
  \includegraphics[width=0.92\textwidth]{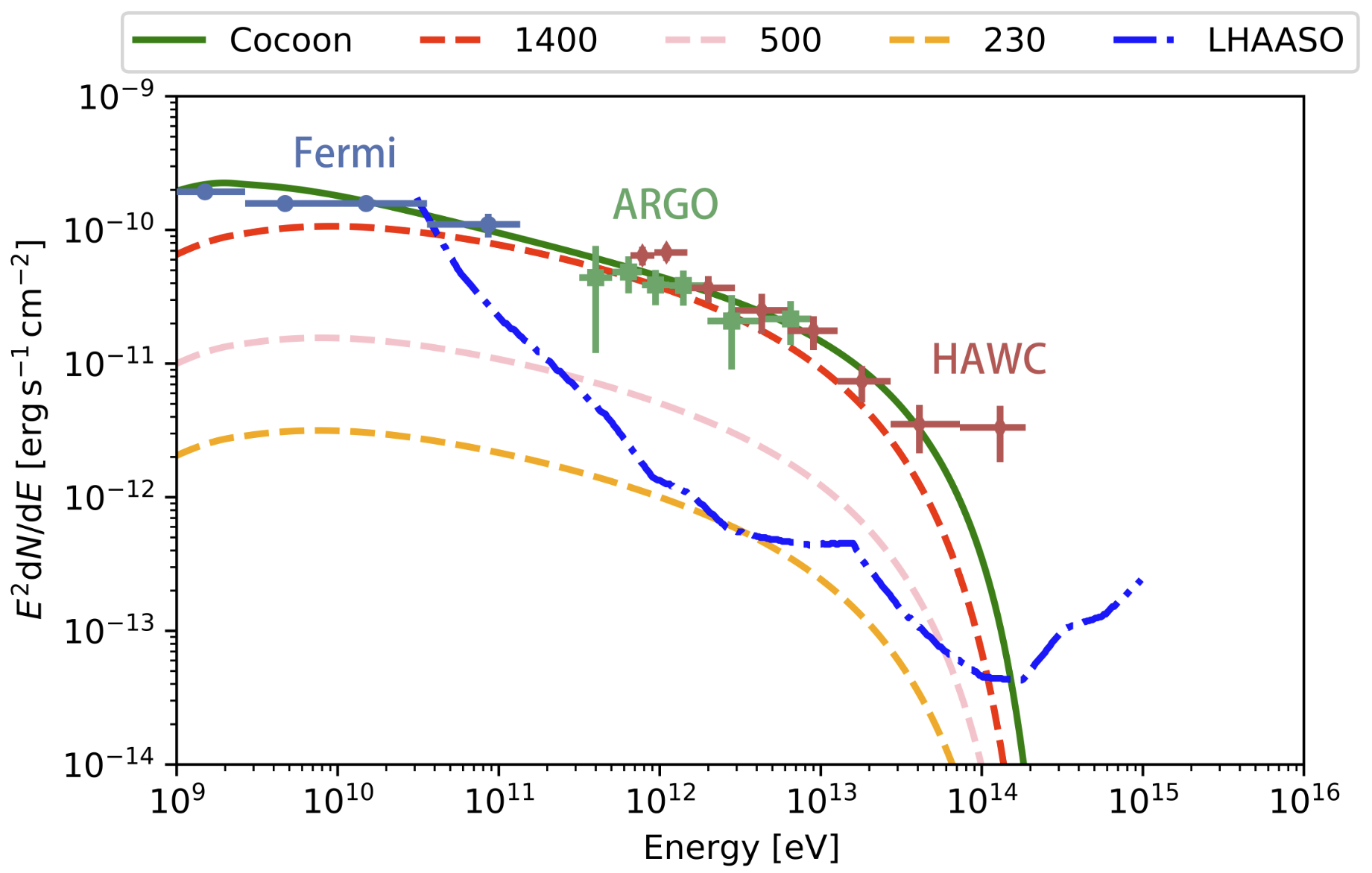}
  \includegraphics[width=0.9\textwidth]{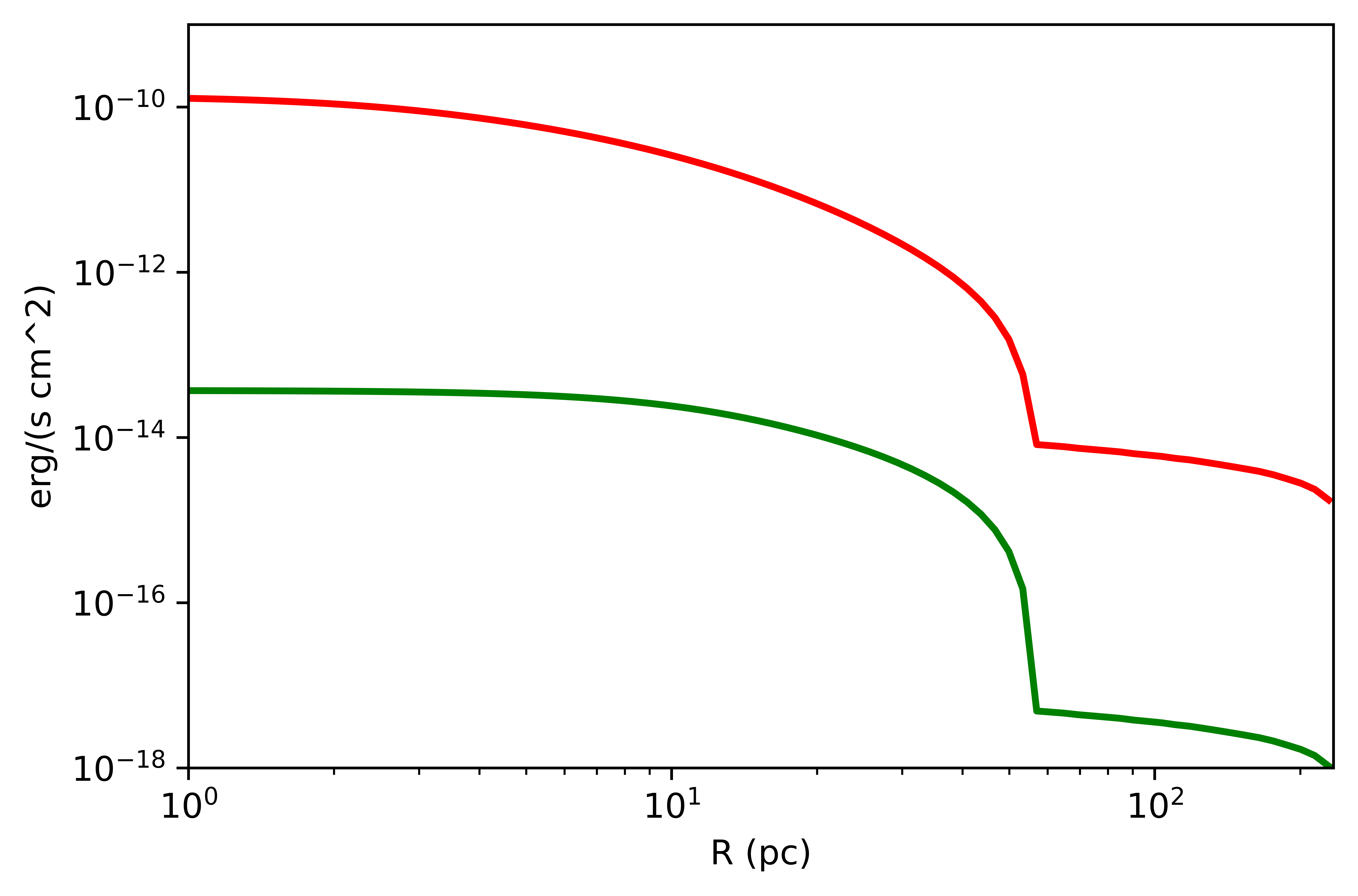}
\end{minipage}
\caption{{\bf Upper panel}: The corresponding protons energy spectrum; the green line represent the protons spectrum in Cygnus Cocoon; the red, pink, and orange dashed lines represent the proton spectra in ISM with 1400 pc, 500 pc, and 230 pc, respectively. {\bf Middle panel}: The estimated differential gamma-ray spectral energy distribution (SED) from Cygnus Cocoon for pure hadronic origin scenario. The green line represent the gamma-ray flux from cocoon inside; the red, pink, and orange dashed lines represent the $\gamma$-ray flux in ISM at 1400 pc, 500 pc, and 230 pc, respectively; the blue chain line represents the differential sensitivity of LHAASO in one year measurement with $20^{\prime \prime}$ PMT in the full operation \citep{2022ChPhC..46c0002C}. {\bf Bottom panel}: The predicted projected radial profile of the gamma-ray flux toward the Cygnus Cocoon. The red line represents the gamma-ray flux in the total energy range 1.4 GeV -- 1 PeV, and the green line represents the flux in the high energy range 50 TeV --1 PeV. }
\label{fig3}
\end{figure}

\begin{table*}
\scriptsize
\caption{The fitted parameter values for the gamma-ray spectrum of Cygnus Cocoon for two cases.}
\begin{tabular}{l c c c c c}
\hline \hline
  & $A(\rm  GeV \, s^{-1})$ & $\alpha$ &  $D_{1}$ ($\rm cm^{2}\, s^{-1}$) ) & $\delta_{1}$ & $E_{\rm max}$ (eV) \\
\hline
Low $E_{\rm max}$ & $6.5\times 10^{38}$ & 2.2 & $1.0\times 10^{25}$  &  0.34 & $3.2\times10^{14}$  \\
High $E_{\rm max}$ & $6.5\times 10^{38}$ & 2.3 & $1.0\times 10^{25}$  &  0.34 & $3.2\times10^{15}$  \\
\hline
\end{tabular}
\label{table1}
\end{table*}

The $\gamma$-ray flux from the Cygnus Cocoon in our model can fit the data from $1 \rm \, GeV$ to $ 50 \rm \, TeV$, but can not explain the hundreds of TeV data. And the model shows that the $\gamma$-ray emission from ISM also has a contribution that cannot be ignored, the $\gamma$-ray flux contribution of ISM with 230 pc range can be detected by LHASSO at $\sim 300\rm\, GeV$,  the $\gamma$-ray flux contribution of ISM with 500 pc range can be detected from $\sim200\rm \, GeV$ to $\sim60 \rm \, TeV$. Then, our model not only predicts that the inside region of the Cygnus Cocoon is bright, but also the ISM surrounding the Cygnus Cocoon might be bright in TeV bands as well. The proton energy spectrum in the Cygnus Cocoon is close to a broken power law with a break at $\sim 1$ TeV, and our model suggests that the distribution of cosmic ray protons in ISM is significantly different compared with the Cygnus Cocoon due to the effect of diffusion. As shown in Table 1, our model gives a very small diffusion constant $\rm D_{1}=1.0\times 10^{25}\rm\, cm^{2}s^{-1}$ compared with that of galactic disk diffusion.

We also try a higher $E_{max}$ to understand the 1.4 PeV photon observed by LHAASO, the rest of the parameters are unchanged. The resulting $\gamma$ -ray and the spectrum of protons are shown in Fig. 4, the relevant parameters are also shown in Table 1. Our model in this case is biased against the data from $15\rm \, TeV$ to $50 \rm \, TeV$, but its flux reaches the sensitivity of LHAASO at $1\rm\, PeV$($\sim 10^{-13}\rm\, erg \, s^{-1}cm^{-2}$), which can explain the $1.4\rm \, PeV$ photon observation. The future observation of the outside region of Cygnus Cocoon around 100 TeV - PeV by LHAASO is the key to constrain our model and diffuse parameters. 

\begin{figure}
\begin{minipage}{0.50\textwidth}
 \includegraphics[width=0.9\textwidth]{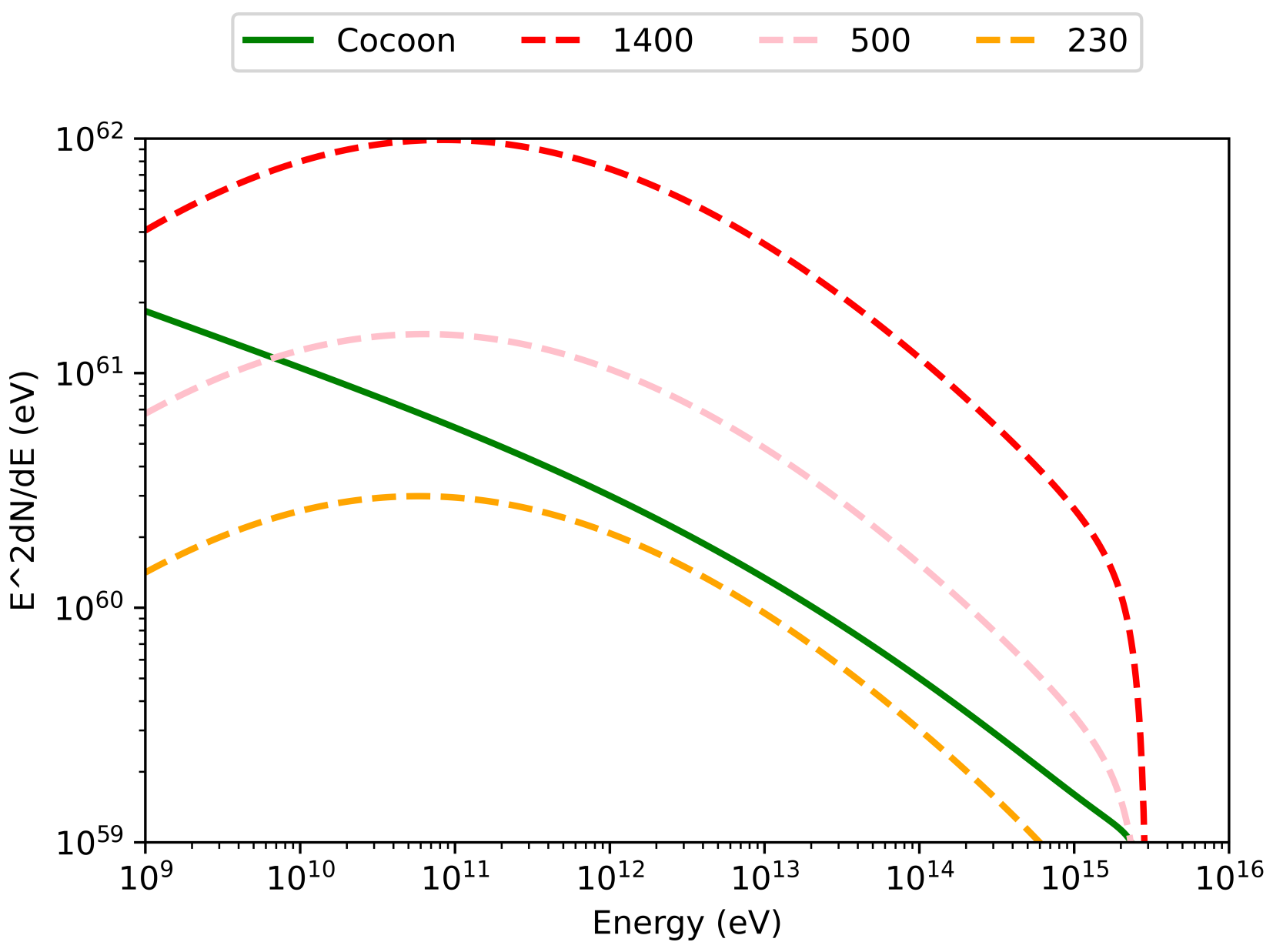}
  \includegraphics[width=0.92\textwidth]{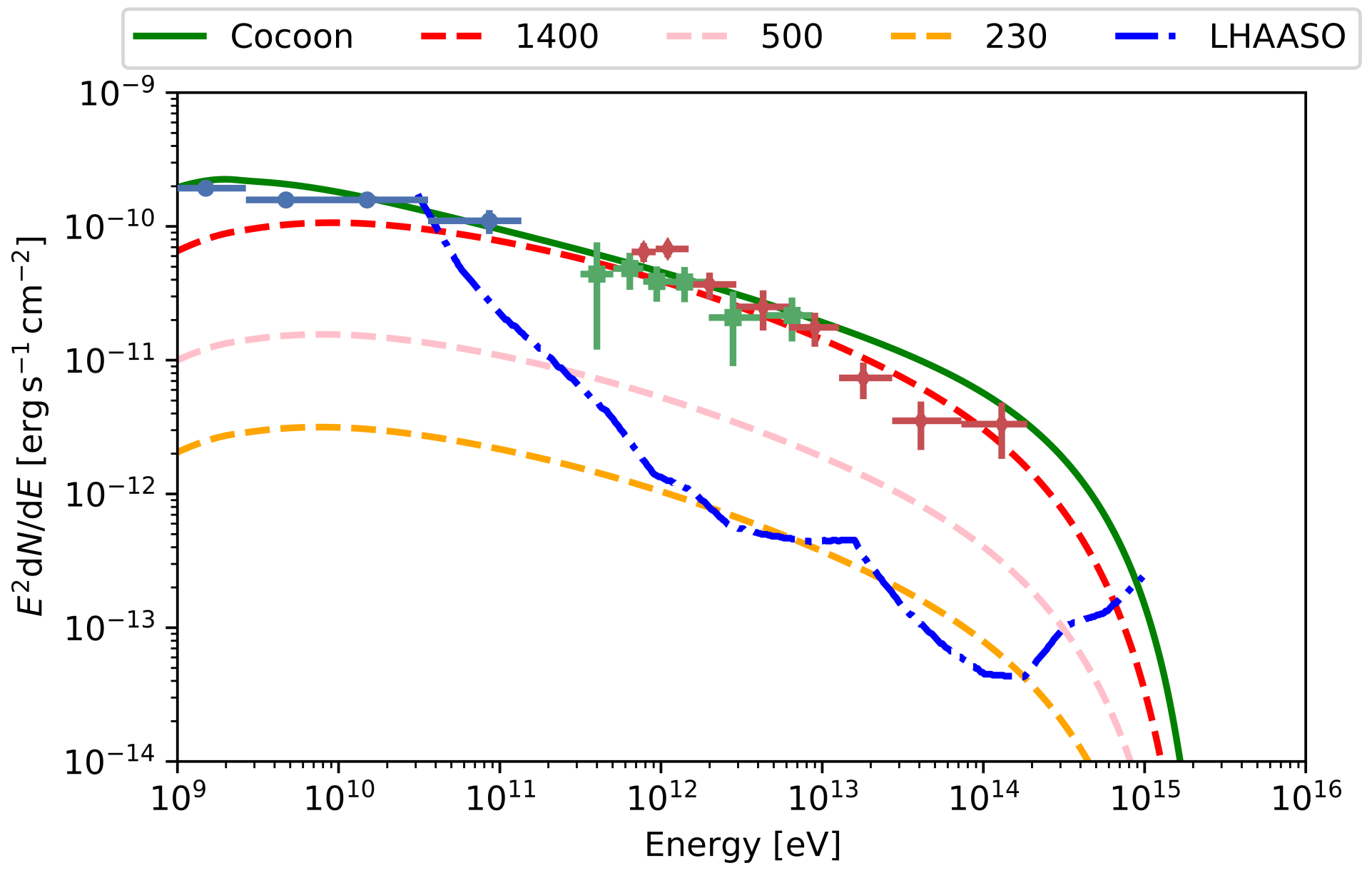}
  \includegraphics[width=0.9\textwidth]{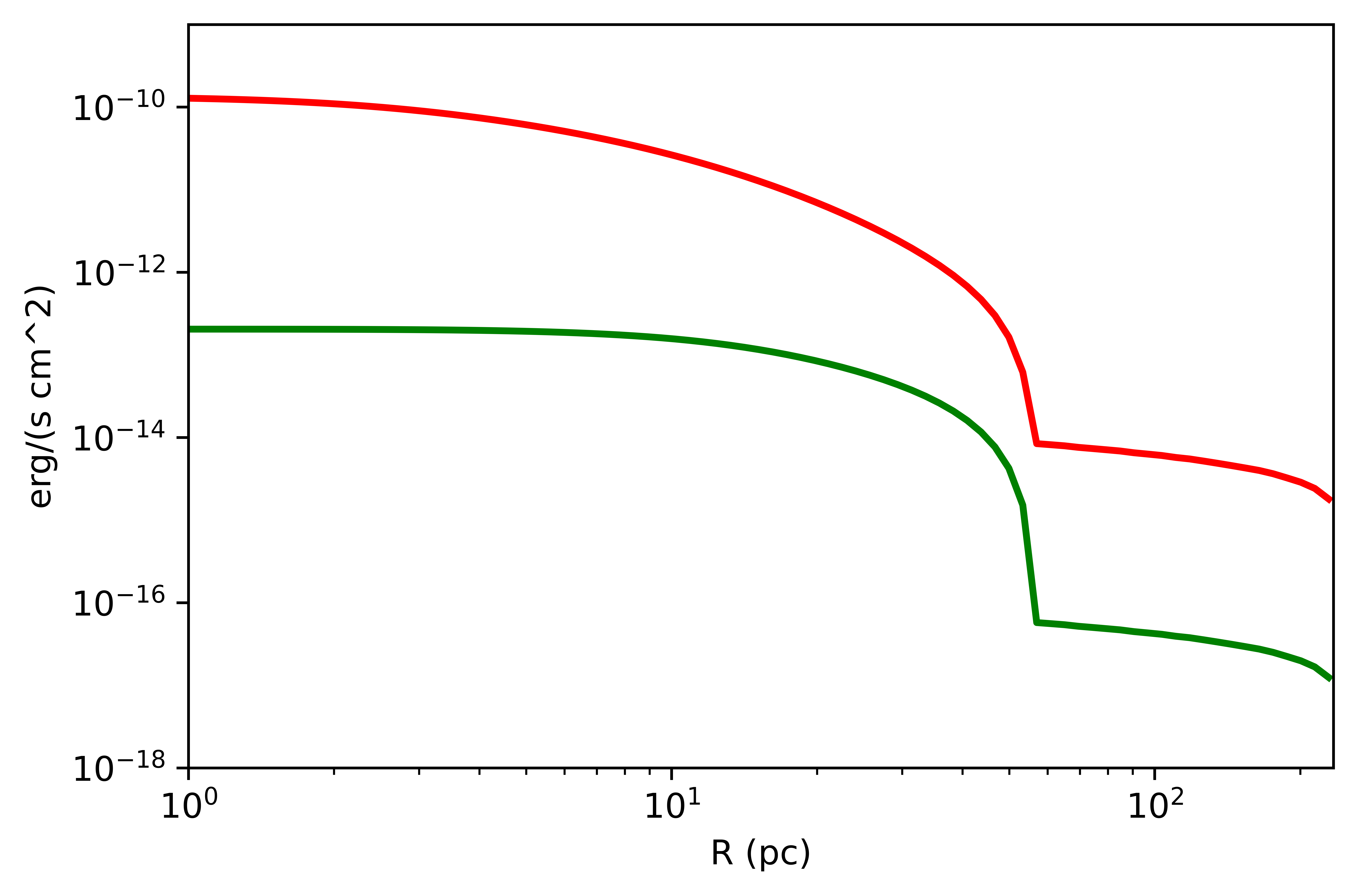} 
\end{minipage}
\caption{The corresponding protons energy spectrum (upper panel), estimated differential gamma-ray spectral energy distribution, and the projected radial profile of the gamma-ray flux (bottom panel), but for a higher $E_{max}$, and other parameters are same with those in Fig. 3.}
\label{fig4}
\end{figure}

Our results show that the $\gamma$-ray emission mainly comes from the Cygnus Cocoon, the $\gamma$-ray flux from ISM within radius $r=230 \rm \, pc$ is only about $1 \%$ or less of the $\gamma$-ray flux from the Cygnus Cocoon at 100 TeV in our model. There are some other possible origins at the sight of the Cygnus Cocoon, such as the Cyg OB2 associations, $\gamma$ Cygni supernova remnant, and a pulsar wind nebula (J2031+415). Although \cite{2021NatAs...5..465A} subtracted the flux from $\gamma$ Cygni and J2031+415, it is still hard to accurately determine what percentage of high energy photons come from the Cygnus Cocoon due to the complex interstellar environment around the Cygnus Cocoon. 

\subsection{Neutrino flux}

The interaction of cosmic-ray protons with ambient matter can produce neutral and charged pions, which can create neutrinos. The number of expected neutrino events in IceCube detector in time $t$ can be calculated by using the following formula:
\begin{equation}
    N_{\nu} = t\int_{E_{\nu,eet}}^{E_{\nu,max}}A_{eff}(E_{\nu})\frac{dn_{\nu}}{dE_{\nu}}dE_{\nu}
\end{equation}
where $E_{\nu,eet}\approx30\rm\, TeV$ is the astrophysical neutrino effective energy threshold of the IceCube detector \citep{2016EPJWC.12604047T},  $A_{eff}$ is the effective area of the Icecube detector which can be given by \cite{2013NuPhS.237..250A}:
\begin{equation}
    A_{eff}(E_{\nu})=16.99\times (\frac{E_{\nu}}{1\rm \, GeV})^{0.2281} \rm \, m^{2}-160.5 \rm \, m^{2}.
\end{equation}
We use the way of \cite{2021PhRvD.104l3027K} to compute the neutrino flux, the results for two cases are shown in Fig. 5. Our model predicted that the neutrino event in 10 yr in lower $E_{max}$ case is estimated to be $N_{\nu}=1.6$ for Cygnus Cocoon, while in higher $E_{max}$ case $N_{\nu}=18.3$ for the Cygnus Cocoon. Our estimate of expected neutrino events in low $E_{max}$ is consistent with the only neutrino event reported by IceCube multiyear observations from the direction of the Cygnus Cocoon so far, but for the high $E_{max}$ case we may predict a more frequent neutrino event report.

\begin{figure}
\begin{minipage}{0.50\textwidth}
 \includegraphics[width=0.88\textwidth]{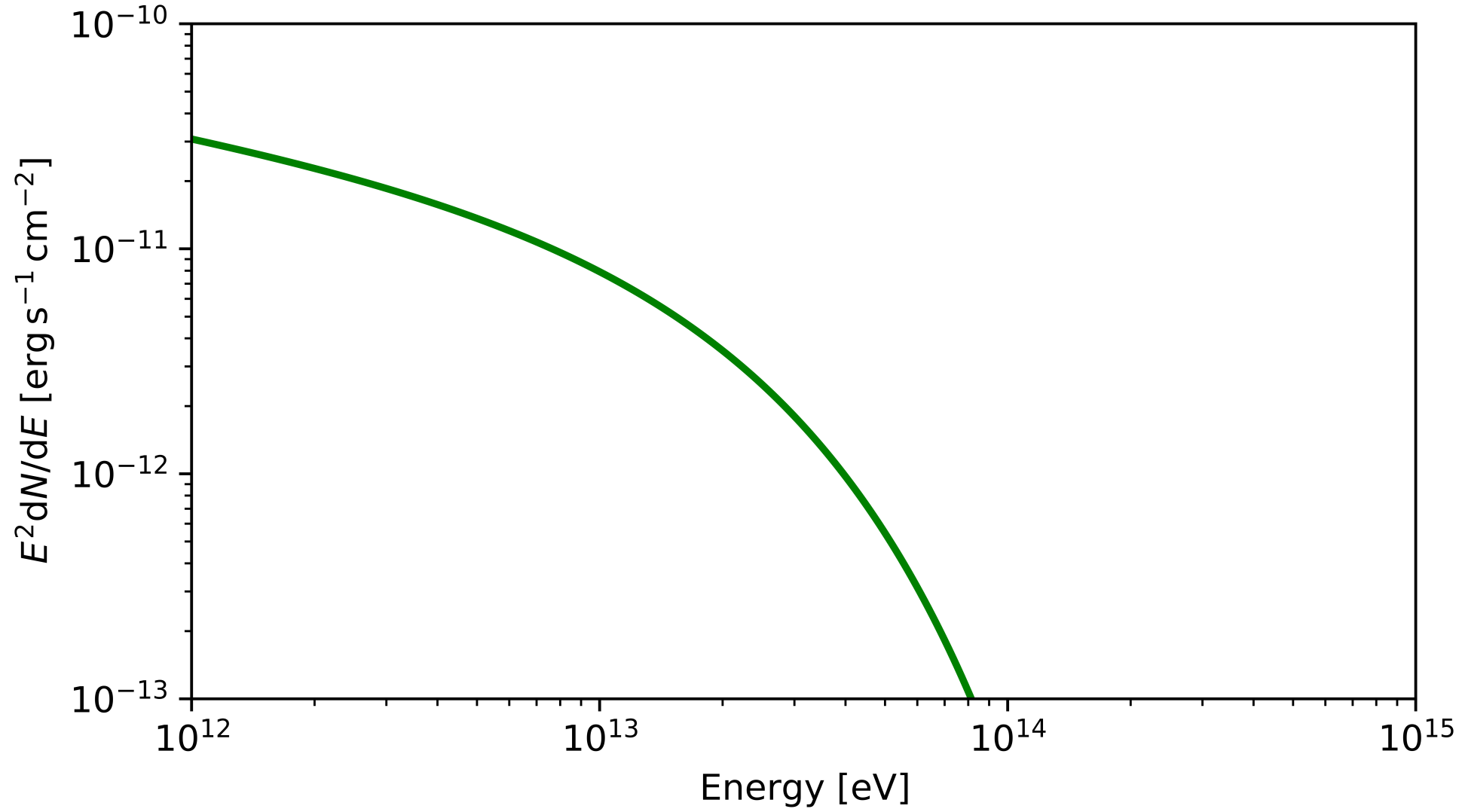}
  \includegraphics[width=0.88\textwidth]{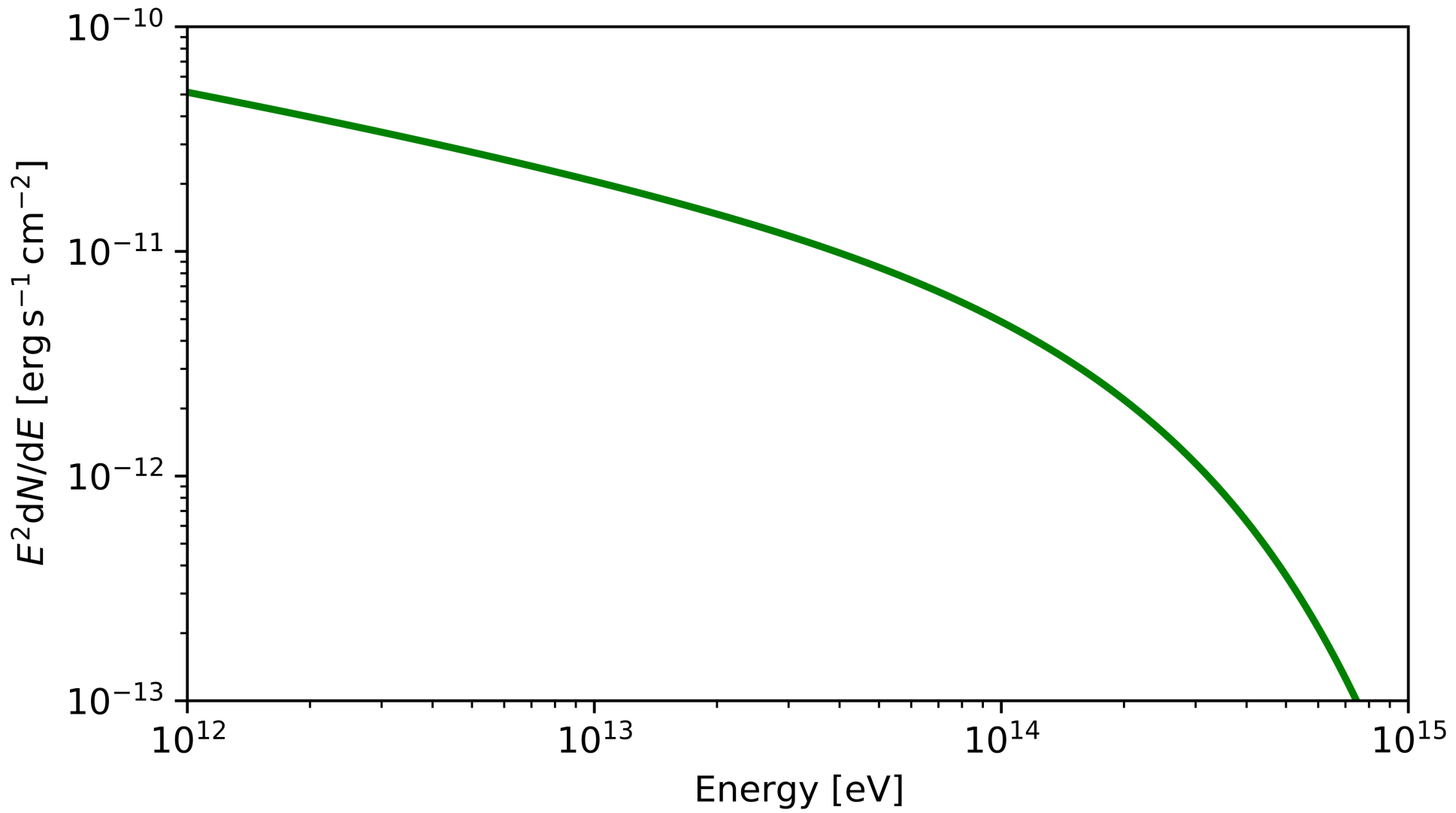}
\end{minipage}
\caption{ {\em \rm Upper panel}: The estimated corresponding neutrino flux. The green line represent the neutrino flux from Cygnus Cocoon. {\em \rm Bottom panel}: Same as the upper panel, but for a higher $E_{max}$. }
\label{fig5}
\end{figure}

\section{Conclusion and Discussions}

The observed gamma rays from the Cygnus Cocoon are difficult to be explained in one single emission site. We discuss a possible model for the high energy particle diffusion  phenomenon for two zones of both inside and outside the Cygnus Cocoon to explain the GeV-PeV spectrum and can well fit the break in gamma-ray spectrum around a few TeVs. Moreover, our model support the possibility that the $1.4 \rm PeV$ photon from the Cygnus OB2 detected LHAASO is produced by the CRs accelerated in the Cygnus Cocoon, although with a higher neutrino frequency compared with the results of IceCube.  

We argue the importance of particle diffusion process in the Cygnus Cocoon. \cite{2022ApJ...931L..30B} discussed the effect of diffusion process on energy spectrum, but the pure hadronic model in their work can not explain the $Feimi$-LAT datas from 10 GeV to 100GeV, on the other side, they add the leptonic process and successfully fit the $\gamma$-ray spectrum for all data except the maximum energy data of HAWC. \cite{2022AdSpR..70.2685B} discuss the effect of plasma mean free path in quasi-resonant magnetic fluctuations which can give the transition energy for CR protons for a pure hadronic model, which also successfully fitted the $\gamma$-ray data. And we use the diffusion–convection equation to calculate the spectrum outside of Cygnus Cocoon in a pure hadronic model, and the diffusion constant $D_1 \sim 10^{25}$ cm$^{2}$ s$^{-1}$ in the Cygnus Cocoon predicted by our model is consistent with the description of \cite{2021NatAs...5..465A}. We also predict that the $\gamma$-ray emission from ISM surrounding the Cygnus Cocoon may also not be ignored. Moreover, our model in both cases can predict the $\gamma$-ray emission from the ISM, which would induce the diffuse gamma-ray structure around the Cygnus cocoon, and can be observed by LHAASO at 100 TeV in near future. The scenario for the high $E_{max}$ predicts a higher frequency neutrino event compared with the present report of IceCube. 

It is not very sure yet that the gamma-ray emission seen by LHAASO is mainly powered by the Cygnus Cocoon. \cite{2021NatAs...5..465A} suggested that pulsar wind nebulae powered by known sources (e.g., PSR J2021+4026, PSR J2032+4127) can not explain the extended Cocoon emission, but the possibility of undiscovered pulsars as the powered sources cannot be ruled out. \cite{2021ApJ...914L...7L} proposed a significant fraction of sub-PeV gamma-rays observed by the Tibet AS+MD array may originate from the source related to the Cygnus region. In future, the better sensitivity gamma-ray telescopes, such as LHAASO (complete operational mode; \citealt{2021Natur.594...33C}), e-ASTROGAM \citep{2018JHEAp..19....1D} may offer a thorough spectral and morphological study for the Cygnus OB2 region. If high-energy $\gamma$-rays up to 100 TeV - PeV in both regions (Cygnus cocoon and outside ISM) are resolved in future, our model could provide a possible physical explanation.

\section*{Acknowledgements}
We are grateful to the referee for the suggestions and Ruoyu Liu for the discussions on the diffusion models. This work is supported by the National Key Research and Development Program of China (Grants No. 2021YFA0718503), the NSFC (12133007, U1838103). 
\section*{Data Availability}
Data that were used in this paper are collected from the published literatures, and available for public users.

\bibliographystyle{mnras}
\bibliography{references}

\label{lastpage}

\end{document}